\documentclass[aps,prl,twocolumn,letterpaper,superscriptaddress,showpacs]{revtex4}
\usepackage{CJK}
\usepackage{keyval}%
\usepackage{graphicx}
\usepackage{dcolumn}
\usepackage{bm}
\usepackage{color}
\usepackage{amsmath}

\setkeys{Gin}{width=0.8\columnwidth,clip=true}

\newcommand{\ignore}[1]{}

\begin{document}
\begin{CJK*}{UTF8}{bsmi}
\title{First-principles method of propagation of tightly bound excitons: exciton band structure of LiF and verification with inelastic x-ray scattering}
\author{Chi-Cheng Lee (%
李啟正
)}
\affiliation{Condensed Matter Physics and Materials Science
Department, Brookhaven National Laboratory, Upton, New York 11973,
USA}%
\affiliation{Institute of Physics, Academia Sinica, Nankang, Taipei 11529, Taiwan}%
\author{Xiaoqian M. Chen (%
\CJKfamily{gbsn}%
陈小千
)}
\affiliation{Department of Physics and Frederick Seitz Materials Research Laboratory, University of Illinois, Urbana, IL 61801, USA}
\author{Yu Gan (%
\CJKfamily{gbsn}%
干禹
)}
\affiliation{Department of Physics and Frederick Seitz Materials Research Laboratory, University of Illinois, Urbana, IL 61801, USA}
\affiliation{Advanced Photon Source, Argonne National Laboratory, Argonne, IL 60439, USA}
\author{Chen-Lin Yeh (%
\CJKfamily{bsmi}%
葉承霖
)}
\affiliation{Condensed Matter Physics and Materials Science
Department, Brookhaven National Laboratory, Upton, New York 11973,
USA}%
\affiliation{Department of Physics, Tamkang University, Tamsui, Taipei 25137, Taiwan}
\author{H. C. Hsueh (%
薛宏中
)}
\affiliation{Department of Physics, Tamkang University, Tamsui, Taipei 25137, Taiwan}
\author{Peter Abbamonte}
\affiliation{Department of Physics and Frederick Seitz Materials Research Laboratory, University of Illinois, Urbana, IL 61801, USA}
\affiliation{Advanced Photon Source, Argonne National Laboratory, Argonne, IL 60439, USA}
\author{Wei Ku (%
顧威
)}
\affiliation{Condensed Matter Physics and Materials Science
Department, Brookhaven National Laboratory, Upton, New York 11973,
USA}%

\date{\today}

\begin{abstract}
We propose a simple first-principles method to describe propagation of tightly bound excitons. 
By viewing the exciton as a composite object (an effective Frenkel exciton in Wannier orbitals), we define 
an exciton kinetic kernel to encapsulate the exciton propagation and decay for all binding energy. 
Applied to prototypical LiF, our approach produces three exciton bands, which we verified quantitatively via inelastic x-ray scattering. 
The proposed real-space picture is computationally inexpensive and thus enables study of the \textit{full} exciton dynamics, 
even in the presence of surfaces and impurity scattering. 
It also provides intuitive understanding to facilitate practical exciton engineering in semiconductors, strongly correlated oxides, 
and their nanostructures. 
\end{abstract}


\maketitle
\end{CJK*}
Excitons, which are elementary excitations comprising a bound electron-hole pair, are crucial in optoelectronic materials applications, such as photovoltaics, solid state lasers, light-emitting diodes, etc.~\cite{Yang,Kanda,Parkinson,Taylor,Hasegawa,Felix,Matsuda}
One of the key issues in such applications is understanding the efficiency with which excitons propagate, as well as their associated decay and recombination, in the diverse contexts in which they arise, including not only bulk semiconductors but also polymers, thin films, quantum dots and various other nanostructured materials~\cite{Yang,Kanda,Parkinson,Taylor,Hasegawa,Felix,Matsuda}.  Recent advances in inelastic x-ray scattering have provided, for the first time, an experimental means to map out the momentum-energy dispersion, and to some extent the internal structure, of excitons~\cite{Peter}.  However this method, like any experiment, yields incomplete information, the cross section being limited by the matrix element of photon-electron coupling.  Understanding the full details of exciton propagation therefore remains a crucial yet challenging task.

Current state-of-the-art theoretical studies of excitons are based either on perturbation theory, via solution to the Bethe-Salpeter equation (BSE)~\cite{Onida2,Rohlfing,Benedict,Hsueh,Onida}, or on specially tuned approximations within the time-dependent density functional theory~\cite{Onida,Bauernschmitt,Reining,Angel}.  While providing great accuracy in some cases, the BSE method requires evaluation of four-point functions with both space and time indices, so are too computationally expensive to describe the full exciton kinetics or to address practical applications.  Application of such methods have therefore been restricted to zero-momentum excitons only, and have not yet broached the issue of exciton propagation.

Furthermore, an increasing number of functional materials now make use of the strongly correlated nature of the electrons, which requires treatment of the many-body problem beyond the framework of perturbation theory.  There is therefore a great need for new, first-principles approaches capable of treating both the propagation and decay of excitions, that can facilitate interpretation of experiments and provide an intuitive, physical picture that deepens our understanding of strongly correlated systems.  Progress in this area would greatly facilitate efforts to tailor exciton properties in functional materials and devices.

In this Letter, we propose a general approach that focuses on the essence of the kinetics of exciton, while allowing various treatments of many-body interactions.
By integrating out the higher-energy virtual pair-breaking processes, the exciton can be treated as a single composite object, instead of a pair of a particle and a hole. Its kinetics can then be formulated in real-space via an exciton kinetic kernel (exciton hopping integral).
This kinetic kernel encapsulates the exciton decay process and the necessary energy dependence to react to all binding energies.
For systems where excitonic effect is large, this real-space description, together with the reduction of the two-body problem to one-body, is significantly more efficient in computation, easily by several orders of magnitude.
This approach can thus account for the \textit{full} exciton kinetics with ease for practical applications.
In addition, the separation of non-local kinetic from the local interacting problem allows an accurate treatment of the local many-body problem beyond perturbation.
Taking the prototypical LiF as a case study, our results give three branches of excitons, which we verified experimentally via inelastic x-ray scattering.
The proposed approach offers a simple and intuitive physical picture easily applicable to systems with surfaces (or other boundary conditions), impurity scattering, or strong many-body interactions.
Therefore, it should facilitate greatly understanding of exciton propagation in all semiconductors, their nanostructures, and numerous strongly correlated functional materials.

The general idea of our approach is to map the generic exciton problem to an ``effective'' Frenkel exciton description, and encapsulate their propagation into an exciton kinetic kernel, like the hopping integral.
This real-space approach makes sense because binding of particles and holes is only significant in short range and thus the size of the exciton with practically relevant binding should be reasonably small.
To this end, one needs to 1) define a ``local'' region, in which the local excitons would be treated accurately enough (beyond perturbation if necessary), and 2) construct the kinetic kernel of the local excitons that describes their propagation and decay in the system.
The first step is strongly case dependent, in both the physical size of the local region and the required treatment of local interaction.
The case study of LiF below gives a detailed example of this step.
The second step, on the other hand, is quite generic and will be the focus of this study.

Let's define the exciton kinetic kernel $T$ through the one-body equation of motion of the exciton propagator $D$ (in matrix notation)
\begin{equation}
\label{eq:eqn1}
D[H] = D[H_L] + D[H_L] T D[H],
\end{equation}
that allows the local exciton (described by $D[H_L]$) to propagate in $D[H]$.
Here $H_L$ refers to the local Hamiltonian, and $H$ the full Hamiltonian containing additional coupling beyond the local region. 
Note that the excitonic degree of freedom $b^\dagger_{\textbf{R}N}$, as a basis for $D_{\textbf{R}N,\textbf{R}^{\prime} N^{\prime}}$ and  $T_{\textbf{R}N,\textbf{R}^{\prime} N^{\prime}}$, is defined in the 
direct product space of the particle $p$ and hole $h$ orbitals at the \textit{same} local region indexed by $\textbf{R}$: $b^\dagger_{\textbf{R}N} = c^\dagger_{\textbf{R}p} c_{\textbf{R}h}$, 
since only strongly bound excitons smaller than the local region are of interest.

This rigorous definition becomes practically useful with the simplification:
\begin{equation}
\label{eq:eqn2}
\begin{aligned}
T \equiv & D^{-1}[H_L] - D^{-1}[H] \\
= & ( D_0^{-1}[G[H_L]] - I[H_L]) - ( D_0^{-1}[G[H]] - I[H]) \\
\cong & D_0^{-1}[G[H_L]] - D_0^{-1}[G[H]].
\end{aligned}
\end{equation}
Here $D_0(\textbf{R}N, \textbf{R}^{\prime} N^{\prime}; t, t^{\prime}) = G( \textbf{R}p, \textbf{R}^{\prime} p^{\prime}; t, t^{\prime}) * G( \textbf{R}^{\prime} h^{\prime}, \textbf{R}h; t^{\prime}, t)$ is the bare polarization bubble describing unbound pairs of fully dressed particles and holes (described by $G$.)
This simplification of the exciton self-energy $I[H] $ by its local counterpart $I[H_L]$ is similar to that previously employed with the dynamical mean field approximation~\cite{Maier}.
It is physically reasonable because of the general short-range nature of the self-energy.
Particularly, at the lowest order approximation of $I$, employed in practically all first-principles implementation of the BSE, $I[H]=I[H_L]$ and this simplification is exact.
Once $T$ is obtained from Eq.~\ref{eq:eqn2}, the full spectrum of $D[H]$ can be obtained effortlessly from Eq.~\ref{eq:eqn1}.

\begin{figure}[tbp]
\includegraphics[width=0.74\columnwidth,clip=true,angle=90]{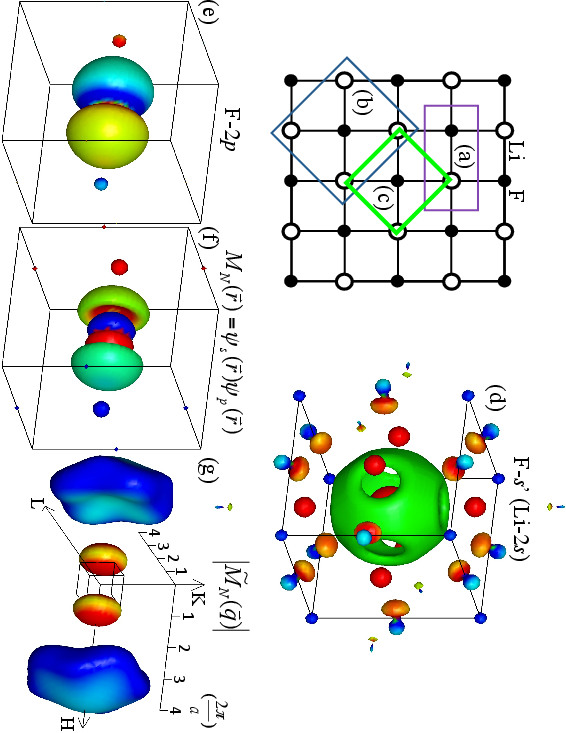}
\caption{\label{fig:fig1} (color online).
Three ways to define local regions via unit cells respecting (a) translational symmetry, (b) rotational symmetry, and (c) both.
In case (c), an effective F-$s^\prime$ Wannier orbital (d) composed of symmetric superposition of Li-2$s$ orbitals is compared to a F-2$p$ Wannier orbital (e).
(f) Shape of the exciton from product of F-$s^\prime$ and F-2$p$ gives the x-ray matrix element $M_N(\textbf q)$ (g) in momentum space.
Boxes in (d), (e), and (f) denote the conventional unit cell.
}
\end{figure}

We now use the prototypical LiF to demonstrate the simple method and the deep insights it can offer.
Crystal LiF has a simple NaCl structure with a 14.2 eV band gap~\cite{Shirley}
defined by three occupied F-2$p$ bands and an unoccupied Li-2$s$ band.
The strong ionicity of the system implies a dominating short-range attraction between a hole in the F-2$p$ orbitals and a particle in the nearest neighboring Li-2$s$ orbital.
We will therefore define a local region containing only one F for simplicity.
The straightforward choice is to include one or six (in 3D) neighboring Li atoms as in Fig.~\ref{fig:fig1}(a) and (b) commonly used in cluster models~\cite{Haverkort}.
These two choices however do not respect the rotational and translational symmetry simultaneously without being overcomplete.
Figure~\ref{fig:fig1}(c) gives a better choice of the local region, containing one F and effectively \textit{one} Li from combining all its neighboring sites.
It respects both the translational and rotational symmetries, and thus form a good basis for the entire lattice.

The low-energy Hilbert space of such a local region can be constructed by making use of the gauge freedom of the Wannier function~\cite{Vanderbilt,Wei}.
Figure~\ref{fig:fig1}(d) shows the resulting Wannier orbital centered at the F site, as a symmetric superposition of atomic Li-2$s$ orbitals.
Recall a similar construction~\cite{Weiguo} in the well-known Zhang-Rice singlet in the cuprate.
Effectively, the local region is just a ``super F atom'' containing three $p$ orbital [Fig.~\ref{fig:fig1}(e)] and one $s^{\prime}$ orbital [Fig.~\ref{fig:fig1}(d)]~\cite{Peter} in a simple fcc lattice, and the tightly bound excitons can now be viewed effectively as intra-``atomic'' Frenkel excitons.
The excitonic Hilbert space $N$ is then simply spanned by three local excitons made of a particle in $p=s^{\prime}$ orbital and a hole in $h=p_x$, $p_y$, or $p_z$ orbital.
Obviously, similar construction applies to all charge transfer insulators~\cite{Lee}.
Fig.~\ref{fig:fig1}(f) also demonstrates the probability amplitude of the exciton in real space, obtained from the product of the particle and hole orbitals~\cite{Larson}.

Now the exciton kinetic kernel can be easily computed via Eq.~\ref{eq:eqn2}.
For a simple illustration, the fully dressed Green's function is obtained approximately from self-consistent DFT Hamiltonian with a +5 eV shift in the on-site energy of $s^{\prime}$ orbital to counter the self-interaction error.
The corresponding band structure contains a 14.2 eV band gap, consistent with the GW calculation~\cite{Shirley,Louie}. 
We then compute the corresponding $D_0[H_L]$ and $D_0[H]$ from the standard convolution~\cite{supplement} in Wannier function basis.
The resulting $D_0[H_L]$ contains trivially a pole at $\omega_0 = \epsilon_p - \epsilon_h =$ 21.9 eV, corresponding to the energy difference between 
the particle $\epsilon_p$ and the hole $\epsilon_h$.
Turning on the local binding interaction $U_B \sim \int dx dx^{\prime} n_s(x) n_p(x^{\prime}) / | x - x^\prime| \sim$ 7 eV simply lowers the pole to $\omega_{exciton} = \omega_0 - U_B =$ 14.9 eV in $D[H_L]$ 
(ignoring the fine multiplet splitting.)
\footnotetext{Our $U_B$ is the binding energy of local bare exciton and is thus much larger than the binding energy between Bloch-like orbitals, typically referred to in the context of Wannier exciton.}

\begin{figure}[tbp]
\includegraphics[width=0.95\columnwidth,clip=true]{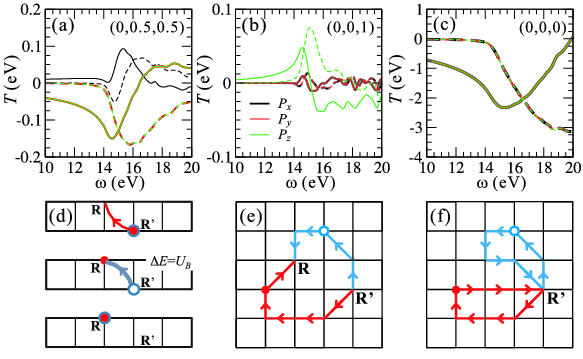}
\caption{\label{fig:fig2} (color online). 
The real (solid lines) and imaginary (dashed lines) parts of exciton kinetic kernel 
$T(\textbf{R}N,\textbf{R}^\prime N)$ for (a) $\textbf{R}^\prime-\textbf{R}$=(0,0.5,0.5), (b) $\textbf{R}^\prime-\textbf{R}$=(0,0,1), 
and (c) $\textbf{R}^\prime-\textbf{R}$=(0,0,0) for $N=P_x$, $P_y$, and $P_z$.
Virtual processes of dressed hopping in strong binding limit (d), unbound particle (red) and hole (blue) motion absorbed in off-site (e) and on-site (f) components of $T$.}
\end{figure}

As shown in Fig.~\ref{fig:fig2}, the resulting exciton kinetic kernel obtained from Eq.~\ref{eq:eqn2} contains several interesting characteristic features.
First, the kinetic kernel is strongly energy dependent.
This is physically necessary because $T$ needs to generate the correct $U_B$-dependent kinetics of excitons \textit{without} the prior knowledge of $U_B$ (to be introduced in $D[H_L]$ through Eq.~\ref{eq:eqn1}.)
Particularly, $T$ hosts a $1/(\omega - \omega_0)\propto 1/U_B$ decay in all its components.
Since $T$ is like the hopping integral, this indicates a heavier mass for excitons with stronger binding.
This is the exact behavior in the strong coupling limit, in which the exciton can only propagate by virtually moving the particle first (with energy $U_B$ in the intermediate state) followed by the hole to reform the pair, a 2nd order process [Fig.~\ref{fig:fig2}(d)] giving $T \sim 2 t_p t_h / U_B \propto 1/U_B$.
($t_p$/$t_h$ refers to the hopping of the particle/hole.)

Second, the kinetic kernel has imaginary part spanning over a wide energy range, representing excitonic decaying channels.
In fact, its energy span is exactly that of the Landau continuum from unbound excitations.
This is not surprising because plugging $D[H_L]$ with unbound excitons ($U_B=0$ and $D[H_L] = D_0[H_L]$) into Eq.~\ref{eq:eqn1} must, by construction, recover exactly the Landau continuum in the resulting $D[H]=D_0[H]$, and the information of the continuum can only come from $T$.
($D_0[H_L]$ is entirely local.)
Physically, this reflects the fact that the definition of $T_{\textbf{R}N,\textbf{R}^{\prime} N^{\prime}}$ via Eq.~\ref{eq:eqn1} integrates out all the virtual processes involving unbound particle and hole traveling independently from $\textbf{R}^{\prime}$ to $\textbf{R}$ [c.f. Fig.~\ref{fig:fig2} (e)].
Therefore, the resulting $T$ describes the exciton kinetics exactly in the weak binding limit as well.

Third, the kinetic kernel has a significant negative diagonal elements [c.f. Fig.~\ref{fig:fig2}(c)] that reduces the average exciton energy to $\widetilde{\omega}_{exciton}\sim$13.4 eV (estimated from $\omega - \omega_{exciton} - \textbf{Re} T_{\textbf{0}N,\textbf{0}N}( \omega) = 0$.)
This renormalization originates from virtual \textit{kinetic} processes illustrated in Fig.~\ref{fig:fig2}(f), and is to be distinguished from the binding of local exciton.


\begin{figure}[tbp]
\includegraphics[width=0.95\columnwidth,clip=true]{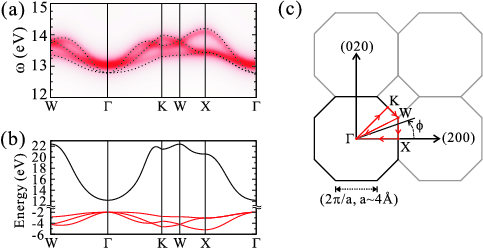}
\caption{\label{fig:fig3} (color online).
(a) Exciton spectrum (red contour) from \textbf{Im}$D(\textbf{q},\omega)$, compared with simple estimation from \textbf{Re}$T(\omega=13.4$ eV$)$ (dotted curves).
(b) LiF band structure containing Li-2$s$ (black) and F-2$p$ (red) bands.
(c) Brillouin zone boundaries and high-symmetry points used in (a) and (b), and selected $k$ paths in the $xy$ plane, denoted by $\phi$, for Fig.~\ref{fig:fig4}.}
\end{figure}

Now, plugging $D[H_L]$ and $T$ into Eq.~\ref{eq:eqn1}, the complete description of exciton propagation and decay is given by the resulting $D[H]$, illustrated by the red spectral intensity of \textbf{Im}$D$ in Fig.~\ref{fig:fig3}(a).
Notice a striking similarity of the dispersion of the exciton bands to that of the red F-$p$ bands of LiF in Fig.~\ref{fig:fig3}(b).
This is because the excitons here have the same symmetry properties as the F-2$p$ orbitals [c.f. Fig.~\ref{fig:fig1}(e) and (f)].
Consequently, $T\sim (2t_p/U_B)t_h$ resembles the hole hopping $t_h$ with a $2t_p/U_B$ reduction that suppresses all longer-range hoppings.
(The comparison appears up-side-down because $t_p < 0$.)
In essence, contrary to the standard effective mass description~\cite{Brus,Vladimir} of Wannier exciton, a strong exciton binding always enhances the tight-binding nature of the excitons, making them propagate more like Frenkel excitons.

The exciton band dispersion can in fact be estimate by simply diagonalizing $\omega_{exciton}+\textbf{Re} T(\widetilde{\omega}_{exciton})$~\cite{supplement}, since $T$ is like the hopping integral for excitons.
Indeed, the the actual dispersion is resembled very closely by the resulting eigenvalue [dotted curves in Fig.~\ref{fig:fig3}(a)], except:
1) a weaker dispersion near the bottom of the exciton bands, due to the above mentioned decay of $T(\omega)$ at lower energy, and
2) broadened exciton peak near the top of the exciton bands, reflecting a shorter lifetime from decaying into the Landau continuum encapsulated in \textbf{Im}$T$.

\begin{figure}[tbp]
\includegraphics[width=0.95\columnwidth,clip=true]{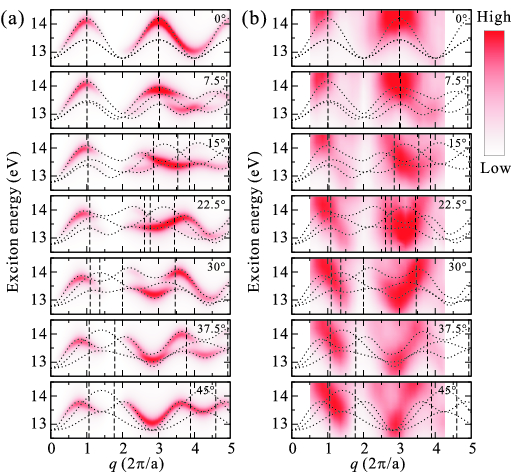}
\caption{\label{fig:fig4} (color online). (a) Imaginary part of dynamical linear response function $\chi$ in selected directions (c.f.:Fig.~\ref{fig:fig3}). The theoretical intensity for $q\leq 1.5$ is reduced by 2.6 to mimic screening by the plasmon pole.
(b) Exciton dispersion measured by inelastic x-ray scattering in the same directions.  The correspondence in the $q-$dependence of the intensity is good, as is visible by comparing the locations of the red regions in panel (a) to the magenta/blue regions in panel (b).  The dispersion and branching are also in good agreement (see the line plots provided in the Supplement~\cite{supplement}).  Vertical dashed lines indicate the Brillouin zone boundaries (c.f.:Fig.~\ref{fig:fig3}).}
\end{figure}

Interestingly, our calcualtion indicates that all three branches of the excitons can be partially observed experimentally.
Naively, given the $p$-character of the probability amplitude $M_N(\textbf q)=\int d\textbf{x}\exp^{-i\textbf{q}\cdot\textbf{x}}\langle \textbf{x}|\textbf{0}p\rangle\langle\textbf{0}h|\textbf{x}\rangle$~\cite{Larson} shown in Fig~\ref{fig:fig1}(g), one expects that only the longitudinal mode can be observed.
However, in reality the zone folding (Umklapp coupling) always hybridizes the longitudinal and the transverse branches near the zone boundaries in most off-high-symmetry directions.
This is clearly demonstrated in our resulting charge susceptability $\chi(\textbf{q},\omega)=\sum_{NN^\prime} M_N(\textbf{q}) D_{NN^\prime}(\textbf{q}) M^*_{N^\prime}(\textbf{q})$, shown in Fig.~\ref{fig:fig4}(a)\footnotemark.
Even with the strong spectral weight suppression in $q < 0.5$, $q\sim 1.5-2$ and $q > 5(2\pi/a)$ from $M_N(\textbf q)$, one still observes a clear switching from one exciton band to another at $q\sim 3.5$ for $\phi=7.5^\circ$, at $q\sim 3.25$ for $30^\circ$, and at $q\sim 3.25$ and $q\sim 4$ for $37.5^\circ$.

\footnotetext{A spectral weight reduction for $q\leq 1.5$ due to plasmon screening is included here in $D$ via additional Hartree contribution~\cite{Lee}.  This improves the agreement in the spectral weight, but has little effects on the exciton dispersion of interest here.}

We verify these results by performing inelastic X-ray scattering along different $q$ directions between (100) and (110)~\cite{supplement}, and comparing to the theoretical response function.
The agreement, illustrated in  Fig.~\ref{fig:fig4}(b), is good in three different respects.
First, there is close correspondence in the momentum dependence of the intensity.
Second, the dispersion agrees well, with close correspondence between the overall bandwidth and locations of the extrema.
Finally, clear indications of the predicted switching between exciton bands (most visible at $\phi=37.5^{\circ}$) are observed, despite the resolution-related broadening in the experimental data.
The dispersion and switching are even more clear when viewed on line plots.~\cite{supplement}
This is, to our knowledge, the first experimental observation of multi-exciton band dispersion in an insulator of any kind.

We stress that the ``invisible'' part of the exciton band structure also represents physical channels of exciton kinetics for propagation, scattering, and decay processes.
It is merely invisible to this particular experimental probe.
Therefore, it is essential to obtain the \textit{full} exciton band structure theoretically.
However, to keep the entire exiton band structure using the standard two-body BSE formulation would be many orders of magnitudes more expensive, if even affordable, as it requires keeping the three momentum- and four band- indices.
In contrast, our approach gives an easy and intuitive access to the full exciton band structure, with superior efficiency necessary for practical applications.

Finally, several important aspects of our method is worth mentioning.
First, our real-space formulation is particularly suitable for treating exciton scattering from surfaces, impurities, or other boundary conditions that disrespect translational symmetry.
Second, the separation of the non-local kinetics from the local binding of exciton in our framework allows a more accurate many-body treatment of the local excitation in $D[H_L]$ (say via exact diagonalization) before applying Eq.~\ref{eq:eqn1}.
This is particularly important in strongly correlated systems, in which significant multiplet splitting~\cite{Lee} exists.
Furthermore, such separation of the problem reduces significantly the computational expense of exciton dynamics:
Evaluation of exciton kinetic kernel $T$ involves no interaction, while the potentially expensive interacting problem is only local.
On top of that, our use of Wannier function as basis, and the implicit absorption of virtual pair-breaking processes further reduce the computation greatly.
For a simple illustration of the superior scaling of our method, the standard BSE requires a $\sim 10^{8}$ elements of the particle-hole interaction~\cite{Louie} , while our method needs only $\sim 10^{4}$ elements. Our real-space approach can incorporate straightforwardly scattering via impurities or boundary conditions, and thus will find broad applications in the studies and engineerings of excitons in modern functional materials, including doped semiconductors and strongly correlated transition metal oxides, as well as their nano-structures, thin films, and heterogeneous interfaces.

Lee thanks useful discussions with T.-K. Lee.
Theoretical work is supported by the U.S. Department of Energy (DOE), Office of Basic Energy Science, under Contract No. DE-AC02-98CH10886 and DOE CMSN. 
IXS measurements are supported by Department of Energy (DOE) grant DE-FG02-06ER46285.
Lee, Yeh, and Hsueh acknowledge support by National Science Council (NSC) and NCTS of Taiwan for support.
Use of the Advanced Photon Source is supported by DOE Contract DE-AC02-O6CH11357.
Use of National Center for High-performance Computing of Taiwan is acknowledged.

* email: weiku@bnl.gov (theory) and abbamonte@mrl.illinois.edu (experiment)

\bibliography{refs}

\begin{thebibliography}{28}
\expandafter\ifx\csname natexlab\endcsname\relax\def\natexlab#1{#1}\fi
\expandafter\ifx\csname bibnamefont\endcsname\relax
  \def\bibnamefont#1{#1}\fi
\expandafter\ifx\csname bibfnamefont\endcsname\relax
  \def\bibfnamefont#1{#1}\fi
\expandafter\ifx\csname citenamefont\endcsname\relax
  \def\citenamefont#1{#1}\fi
\expandafter\ifx\csname url\endcsname\relax
  \def\url#1{\texttt{#1}}\fi
\expandafter\ifx\csname urlprefix\endcsname\relax\def\urlprefix{URL }\fi
\providecommand{\bibinfo}[2]{#2}
\providecommand{\eprint}[2][]{\url{#2}}

\bibitem[{\citenamefont{Huang et~al.}(2001)\citenamefont{Huang, Mao, Feick,
  Yan, Wu, Kind, Weber, Russo, and Yang}}]{Yang}
\bibinfo{author}{\bibfnamefont{M.~H.} \bibnamefont{Huang}},
  \bibinfo{author}{\bibfnamefont{S.}~\bibnamefont{Mao}},
  \bibinfo{author}{\bibfnamefont{H.}~\bibnamefont{Feick}},
  \bibinfo{author}{\bibfnamefont{H.}~\bibnamefont{Yan}},
  \bibinfo{author}{\bibfnamefont{Y.}~\bibnamefont{Wu}},
  \bibinfo{author}{\bibfnamefont{H.}~\bibnamefont{Kind}},
  \bibinfo{author}{\bibfnamefont{E.}~\bibnamefont{Weber}},
  \bibinfo{author}{\bibfnamefont{R.}~\bibnamefont{Russo}}, \bibnamefont{and}
  \bibinfo{author}{\bibfnamefont{R.}~\bibnamefont{Yang}},
  \bibinfo{journal}{Science} \textbf{\bibinfo{volume}{292}},
  \bibinfo{pages}{1897} (\bibinfo{year}{2001}).

\bibitem[{\citenamefont{Koizumi et~al.}(2001)\citenamefont{Koizumi, Watanabe,
  Hasegawa, and Kanda}}]{Kanda}
\bibinfo{author}{\bibfnamefont{S.}~\bibnamefont{Koizumi}},
  \bibinfo{author}{\bibfnamefont{K.}~\bibnamefont{Watanabe}},
  \bibinfo{author}{\bibfnamefont{M.}~\bibnamefont{Hasegawa}}, \bibnamefont{and}
  \bibinfo{author}{\bibfnamefont{H.}~\bibnamefont{Kanda}},
  \bibinfo{journal}{Science} \textbf{\bibinfo{volume}{292}},
  \bibinfo{pages}{1899} (\bibinfo{year}{2001}).

\bibitem[{\citenamefont{Sambur et~al.}(2010)\citenamefont{Sambur, Novet, and
  Parkinson}}]{Parkinson}
\bibinfo{author}{\bibfnamefont{J.~B.} \bibnamefont{Sambur}},
  \bibinfo{author}{\bibfnamefont{T.}~\bibnamefont{Novet}}, \bibnamefont{and}
  \bibinfo{author}{\bibfnamefont{B.~A.} \bibnamefont{Parkinson}},
  \bibinfo{journal}{Science} \textbf{\bibinfo{volume}{330}},
  \bibinfo{pages}{63} (\bibinfo{year}{2010}).

\bibitem[{\citenamefont{Thorsm{\o}lle et~al.}(2009)}]{Taylor}
\bibinfo{author}{\bibfnamefont{V.~K.} \bibnamefont{Thorsm{\o}lle}}
  \bibnamefont{et~al.}, \bibinfo{journal}{Phys. Rev. Lett.}
  \textbf{\bibinfo{volume}{102}}, \bibinfo{pages}{017401}
  (\bibinfo{year}{2009}).

\bibitem[{\citenamefont{Tsutsumi et~al.}(2010)}]{Hasegawa}
\bibinfo{author}{\bibfnamefont{J.}~\bibnamefont{Tsutsumi}}
  \bibnamefont{et~al.}, \bibinfo{journal}{Phys. Rev. Lett.}
  \textbf{\bibinfo{volume}{105}}, \bibinfo{pages}{226601}
  (\bibinfo{year}{2010}).

\bibitem[{\citenamefont{Deschler et~al.}(2011)}]{Felix}
\bibinfo{author}{\bibfnamefont{F.}~\bibnamefont{Deschler}}
  \bibnamefont{et~al.}, \bibinfo{journal}{Phys. Rev. Lett.}
  \textbf{\bibinfo{volume}{107}}, \bibinfo{pages}{127402}
  (\bibinfo{year}{2011}).

\bibitem[{\citenamefont{Matsuda et~al.}(2003)\citenamefont{Matsuda, Saiki,
  Nomura, Mihara, Aoyagi, Nair, and Takagahara}}]{Matsuda}
\bibinfo{author}{\bibfnamefont{K.}~\bibnamefont{Matsuda}},
  \bibinfo{author}{\bibfnamefont{T.}~\bibnamefont{Saiki}},
  \bibinfo{author}{\bibfnamefont{S.}~\bibnamefont{Nomura}},
  \bibinfo{author}{\bibfnamefont{M.}~\bibnamefont{Mihara}},
  \bibinfo{author}{\bibfnamefont{Y.}~\bibnamefont{Aoyagi}},
  \bibinfo{author}{\bibfnamefont{S.}~\bibnamefont{Nair}}, \bibnamefont{and}
  \bibinfo{author}{\bibfnamefont{T.}~\bibnamefont{Takagahara}},
  \bibinfo{journal}{Phys. Rev. Lett.} \textbf{\bibinfo{volume}{91}},
  \bibinfo{pages}{177401} (\bibinfo{year}{2003}).

\bibitem[{\citenamefont{Abbamonte et~al.}(2008)}]{Peter}
\bibinfo{author}{\bibfnamefont{P.}~\bibnamefont{Abbamonte}}
  \bibnamefont{et~al.}, \bibinfo{journal}{PNAS} \textbf{\bibinfo{volume}{105}},
  \bibinfo{pages}{12159} (\bibinfo{year}{2008}).

\bibitem[{\citenamefont{Onida et~al.}(1995)}]{Onida2}
\bibinfo{author}{\bibfnamefont{G.}~\bibnamefont{Onida}} \bibnamefont{et~al.},
  \bibinfo{journal}{Phys. Rev. Lett.} \textbf{\bibinfo{volume}{75}},
  \bibinfo{pages}{818} (\bibinfo{year}{1995}).

\bibitem[{\citenamefont{Rohlfing and Louie}(1998)}]{Rohlfing}
\bibinfo{author}{\bibfnamefont{M.}~\bibnamefont{Rohlfing}} \bibnamefont{and}
  \bibinfo{author}{\bibfnamefont{S.~G.} \bibnamefont{Louie}},
  \bibinfo{journal}{Phys. Rev. Lett.} \textbf{\bibinfo{volume}{81}},
  \bibinfo{pages}{2312} (\bibinfo{year}{1998}).

\bibitem[{\citenamefont{Benedict et~al.}(1998)}]{Benedict}
\bibinfo{author}{\bibfnamefont{L.~X.} \bibnamefont{Benedict}}
  \bibnamefont{et~al.}, \bibinfo{journal}{Phys. Rev. Lett.}
  \textbf{\bibinfo{volume}{80}}, \bibinfo{pages}{4514} (\bibinfo{year}{1998}).

\bibitem[{\citenamefont{Hsueh et~al.}(2011)\citenamefont{Hsueh, Guo, and
  Louie}}]{Hsueh}
\bibinfo{author}{\bibfnamefont{H.~C.} \bibnamefont{Hsueh}},
  \bibinfo{author}{\bibfnamefont{G.~Y.} \bibnamefont{Guo}}, \bibnamefont{and}
  \bibinfo{author}{\bibfnamefont{S.~G.} \bibnamefont{Louie}},
  \bibinfo{journal}{Phys. Rev. B} \textbf{\bibinfo{volume}{84}},
  \bibinfo{pages}{085404} (\bibinfo{year}{2011}).

\bibitem[{\citenamefont{Onida et~al.}(2002)}]{Onida}
\bibinfo{author}{\bibfnamefont{G.}~\bibnamefont{Onida}} \bibnamefont{et~al.},
  \bibinfo{journal}{Rev. Mod. Phys.} \textbf{\bibinfo{volume}{74}},
  \bibinfo{pages}{601} (\bibinfo{year}{2002}).

\bibitem[{\citenamefont{Bauernschmitt and Ahlrichs}(1996)}]{Bauernschmitt}
\bibinfo{author}{\bibfnamefont{R.}~\bibnamefont{Bauernschmitt}}
  \bibnamefont{and} \bibinfo{author}{\bibfnamefont{R.}~\bibnamefont{Ahlrichs}},
  \bibinfo{journal}{Chem. Phys. Lett.} \textbf{\bibinfo{volume}{256}},
  \bibinfo{pages}{454} (\bibinfo{year}{1996}).

\bibitem[{\citenamefont{Reining et~al.}(2002)}]{Reining}
\bibinfo{author}{\bibfnamefont{L.}~\bibnamefont{Reining}} \bibnamefont{et~al.},
  \bibinfo{journal}{Phys. Rev. Lett.} \textbf{\bibinfo{volume}{88}},
  \bibinfo{pages}{066404} (\bibinfo{year}{2002}).

\bibitem[{\citenamefont{Marini et~al.}(2003)\citenamefont{Marini, Sole, and
  Rubio}}]{Angel}
\bibinfo{author}{\bibfnamefont{A.}~\bibnamefont{Marini}},
  \bibinfo{author}{\bibfnamefont{R.~D.} \bibnamefont{Sole}}, \bibnamefont{and}
  \bibinfo{author}{\bibfnamefont{A.}~\bibnamefont{Rubio}},
  \bibinfo{journal}{Phys. Rev. Lett.} \textbf{\bibinfo{volume}{91}},
  \bibinfo{pages}{256402} (\bibinfo{year}{2003}).

\bibitem[{\citenamefont{Maier et~al.}(2000)\citenamefont{Maier, Jarrell,
  Pruschke, and Keller}}]{Maier}
\bibinfo{author}{\bibfnamefont{T.}~\bibnamefont{Maier}},
  \bibinfo{author}{\bibfnamefont{M.}~\bibnamefont{Jarrell}},
  \bibinfo{author}{\bibfnamefont{T.}~\bibnamefont{Pruschke}}, \bibnamefont{and}
  \bibinfo{author}{\bibfnamefont{J.}~\bibnamefont{Keller}},
  \bibinfo{journal}{Phys. Rev. Lett.} \textbf{\bibinfo{volume}{85}},
  \bibinfo{pages}{1524} (\bibinfo{year}{2000}).

\bibitem[{\citenamefont{Shirley et~al.}(1996)\citenamefont{Shirley, Terminello,
  Klepeis, and Himpsel}}]{Shirley}
\bibinfo{author}{\bibfnamefont{E.~L.} \bibnamefont{Shirley}},
  \bibinfo{author}{\bibfnamefont{L.~J.} \bibnamefont{Terminello}},
  \bibinfo{author}{\bibfnamefont{J.~E.} \bibnamefont{Klepeis}},
  \bibnamefont{and} \bibinfo{author}{\bibfnamefont{F.~J.}
  \bibnamefont{Himpsel}}, \bibinfo{journal}{Phys. Rev. B}
  \textbf{\bibinfo{volume}{53}}, \bibinfo{pages}{10296} (\bibinfo{year}{1996}),
  \bibinfo{note}{and references therein}.

\bibitem[{\citenamefont{Haverkort et~al.}(2007)}]{Haverkort}
\bibinfo{author}{\bibfnamefont{M.~W.} \bibnamefont{Haverkort}}
  \bibnamefont{et~al.}, \bibinfo{journal}{Phys. Rev. Lett.}
  \textbf{\bibinfo{volume}{99}}, \bibinfo{pages}{257401}
  (\bibinfo{year}{2007}).

\bibitem[{\citenamefont{Marzari and Vanderbilt}(1997)}]{Vanderbilt}
\bibinfo{author}{\bibfnamefont{N.}~\bibnamefont{Marzari}} \bibnamefont{and}
  \bibinfo{author}{\bibfnamefont{D.}~\bibnamefont{Vanderbilt}},
  \bibinfo{journal}{Phys. Rev. B} \textbf{\bibinfo{volume}{56}},
  \bibinfo{pages}{12847} (\bibinfo{year}{1997}).

\bibitem[{\citenamefont{Ku et~al.}(2002)}]{Wei}
\bibinfo{author}{\bibfnamefont{W.}~\bibnamefont{Ku}} \bibnamefont{et~al.},
  \bibinfo{journal}{Phys. Rev. Lett.} \textbf{\bibinfo{volume}{89}},
  \bibinfo{pages}{167204} (\bibinfo{year}{2002}).

\bibitem[{\citenamefont{Yin and Ku}(2009)}]{Weiguo}
\bibinfo{author}{\bibfnamefont{W.-G.} \bibnamefont{Yin}} \bibnamefont{and}
  \bibinfo{author}{\bibfnamefont{W.}~\bibnamefont{Ku}}, \bibinfo{journal}{Phys.
  Rev. B} \textbf{\bibinfo{volume}{79}}, \bibinfo{pages}{214512}
  (\bibinfo{year}{2009}).

\bibitem[{\citenamefont{Lee et~al.}(2010)\citenamefont{Lee, Hsueh, and
  Ku}}]{Lee}
\bibinfo{author}{\bibfnamefont{C.-C.} \bibnamefont{Lee}},
  \bibinfo{author}{\bibfnamefont{H.~C.} \bibnamefont{Hsueh}}, \bibnamefont{and}
  \bibinfo{author}{\bibfnamefont{W.}~\bibnamefont{Ku}}, \bibinfo{journal}{Phys.
  Rev. B} \textbf{\bibinfo{volume}{82}}, \bibinfo{pages}{081106}
  (\bibinfo{year}{2010}).

\bibitem[{\citenamefont{Larson et~al.}(2007)}]{Larson}
\bibinfo{author}{\bibfnamefont{B.~C.} \bibnamefont{Larson}}
  \bibnamefont{et~al.}, \bibinfo{journal}{Phys. Rev. Lett.}
  \textbf{\bibinfo{volume}{99}}, \bibinfo{pages}{026401}
  (\bibinfo{year}{2007}).

\bibitem[{\citenamefont{Rohlfing and Louie}(2000)}]{Louie}
\bibinfo{author}{\bibfnamefont{M.}~\bibnamefont{Rohlfing}} \bibnamefont{and}
  \bibinfo{author}{\bibfnamefont{S.~G.} \bibnamefont{Louie}},
  \bibinfo{journal}{Phys. Rev. B} \textbf{\bibinfo{volume}{62}},
  \bibinfo{pages}{4927} (\bibinfo{year}{2000}).

\bibitem[{sup()}]{supplement}
\eprint{See EPAPS Document No. for the details of IXS experiment and
  computational parameters.}

\bibitem[{\citenamefont{Brus}(1984)}]{Brus}
\bibinfo{author}{\bibfnamefont{L.~E.} \bibnamefont{Brus}}, \bibinfo{journal}{J.
  Chem. Phys.} \textbf{\bibinfo{volume}{80}}, \bibinfo{pages}{4403}
  (\bibinfo{year}{1984}).

\bibitem[{\citenamefont{Agranovich}(2009)}]{Vladimir}
\bibinfo{author}{\bibfnamefont{V.~M.} \bibnamefont{Agranovich}},
  \emph{\bibinfo{title}{Excitations in Organic Solids}}
  (\bibinfo{publisher}{Oxford University Press}, \bibinfo{year}{2009}).

\end{thebibliography}
\end{document}